# Stock Forecasting using M-Band Wavelet-Based SVR and RNN-LSTMs Models

Hieu Quang Nguyen*
*Department of Mathematics*
*Western Connecticut State University*
Danbury, CT, 06810, U.S.A.
nguyen084@connect.wcsu.edu

Abdul Hasib Rahimyar*
*Department of Mathematics*
*Western Connecticut State University*
Danbury, CT, 06810, U.S.A.
rahimyar001@connect.wcsu.edu

Xiaodi Wang, PhD
*Department of Mathematics*
*Western Connecticut State University*
Danbury, CT, 06810, U.S.A.
wangx@wcsu.edu

*Abstract*—The task of predicting future stock values has always been one that is heavily desired albeit very difficult. This difficulty arises from stocks with non-stationary behavior, and without any explicit form. Hence, predictions are best made through analysis of financial stock data. To handle big data sets, current convention involves the use of the Moving Average. However, by utilizing the Wavelet Transform in place of the Moving Average to denoise stock signals, financial data can be smoothened and more accurately broken down. This newly transformed, denoised, and more stable stock data can be followed up by non-parametric statistical methods, such as Support Vector Regression (SVR) and Recurrent Neural Network (RNN) based Long Short-Term Memory (LSTM) networks to predict future stock prices. Through the implementation of these methods, one is left with a more accurate stock forecast, and in turn, increased profits.

## I. INTRODUCTION

The forecasting of the financial market is an appealing albeit challenging task. It has been an attractive research topic for many practitioners, scholars and researchers from various academic fields. To achieve an accurate result, one must utilize historical data and apply better prediction algorithms. The management and analysis of financial data are thus imperative to accurately forecasting the future price of a given stock.

This methodology can be extended into the field of options trading; a form of investing where the investor has the right to buy or sell a stock at an agreed-upon price over an agreed upon period of time. Clearly, being able to predict the future price of a stock holds immense importance in how successful an option will be.

However, the handling of financial data is a demanding task in and of itself. Any given stock is associated with high volatility, a small sample size, non-stationary and non-linearity behavior, and numerous high noise variables, all of which make its analysis difficult. High noise data leads the incomplete information gap between a stock's historical trading price and its future price. Currently, when it comes to handling and denoising financial data, the industry standard is to use what is known as a moving average. Although moving averages have been shown to be a feasible way of handling data, we propose a replacement through the Wavelet Denoise Procedure that will yield improved results.

The Wavelet Transform is a data analysis technique in which a dataset is transformed to the Wavelet domain and then broken down into low and high-frequency portions. It is believed that most of the data's noise is hiding in the high-frequency portion of its Wavelet domain. Through the denoising procedure, one can isolate and remove much of the undesired noise that is present in almost every stock, all of which contribute to a stock's volatility but is not necessarily indicative of the stock's performance. After the application of the Wavelet Transform and the denoising procedure, one is left with a data set that better represents the stock's true behavior and that is void of much of its external noise; all of which is crucial in achieving a more accurate stock forecast. The theory and application of the Wavelet Transform are discussed in further detail in Sections C, Chapter III.

To assess the accuracy of this method, each denoised dataset was followed up with various Machine Learning techniques, namely Support Vector Regression (SVR) and Recurrent Neural Network (RNN) with Long Short-Term Memory (LSTM) networks; all of which can be used to predict the movement of a stock. More detail behind each technique can be found in Sections A and B from Chapter III, and their corresponding methodologies are outlined in Chapter IV. The results from this study can be found in Sections B and C, Chapter IV.

## II. LITERATURE REVIEW

In recent years, accurately forecasting the prices of financial products/indices has become an ever-increasingly important issue in investment decision making. Traditionally, many prediction models have focused on linear statistical time series models such as ARIMA [24]. However, financial time series are inherently noisy, evolutionary, and non-linear dynamical systems [1]. Neural Networks (NNs) are useful techniques for modeling financial time series due to their ability to capture hidden functional relationships within financial data [1][17-20]. Recently, substantial criticism of NNs has found the vanishing gradient problem of traditional NNs, and raised skepticism regarding their ability to forecast even simple time series patterns of seasonality or trends without prior data preprocessing [37]. To address these drawbacks, numerous attempts have been made. In [15][25-26], Wavelet Transforms are used in the regressions without any assumptions regarding the form of the underlying functions, which possess various superior qualities that are illustrated through their actual performances in forecasting S&P 500 index and some variance functions. On the other hand, [2-3][12-13][21][23] used Recurrent Neural Networks (RNNs) based on Long Short-Term Memory (LSTM) networks to overcome shortcomings of traditional RNNs. At the same time, Support Vector Regression (SVR) has also become increasingly popular as a nonparametric tool to be applied in the predicting of financial time series data [32]. Perhaps the best feature of SVR is that the capacity of the system is controlled by parameters that are not dependent on the dimensionality of the feature space. There is always an effort to seek and optimize the generalization bounds given for regression. A loss function that ignores errors and is situated within a certain distance of the true value is used for the purpose of regression [4-5][8][22][24][28-29]. SVR has evolved in the framework of statistical learning theory [29] and can be utilized for problems involving linear or nonlinear regression. Another significant characteristic is that it involves the solution of a convex optimization problem with a unique global (and also sparse) solution. In contrast, other nonparametric methods used for regression analysis usually have non-convex error functions that entail the risk of having multiple local minima solutions [5][22][29].

In this paper, our main contribution is to develop Wavelet-based Machine Learning models for forecasting stock price/return. We apply SVR with different kernels and LSTMs to the M-band Wavelet Transformed and denoised data.

III. PRIMARIES

A. Support Vector Regression (SVR)

SVR is a non-parametric kernel-based regression method used for extrapolating future values. More specifically, we shall be focusing on ε-SVR, a form of SVR in which a hyperplane is constructed with a loss function within ε precision. The SVR function can be expressed as [29][34]:

$$f(x) = w^T \varphi(x) + b \quad (1)$$

where $\varphi(x)$ maps data from the input space to the feature space, $w$ is a weight vector, and $b$ is a bias constant. $w$ and $b$ are estimated by satisfying the following [29][34]:

Minimizing: $\frac{1}{2}\|w\|^2$
Subject to:

$$\begin{cases} y_i - (\langle w, \varphi(x_i)\rangle + b) \leq \varepsilon \\ (\langle w, \varphi(x_i)\rangle + b) - y_i \leq \varepsilon \end{cases} \quad (2)$$

where $x_i$ and $y_i$ represent input/target values obtained from the training set. To address points outside this ε-insensitive band, we introduce slack variables $\xi_i, \xi_i^*$ [29][34]:

Minimizing: $\frac{1}{2}\|w\|^2 + C\sum_{i=1}^{l}(\xi_i + \xi_i^*)$
Subject to:

$$\begin{cases} y_i - (\langle w, \varphi(x_i)\rangle + b) \leq \varepsilon + \xi_i \\ (\langle w, \varphi(x_i)\rangle + b) - y_i \leq \varepsilon + \xi_i^* \\ \xi_i, \xi_i^* \geq 0 \end{cases} \quad (3)$$

The constant $C > 0$ is used to represent the tradeoff between model complexity and training error [29][34]. After taking the Lagrangian and optimizing with the above constraints, we are left with [29][34]:

$$f(x) = \sum_{i=1}^{l}(\alpha_i - \alpha_i^*)K(x_i, x) + b \quad (4)$$

where $K(\cdot,\cdot)$ is a kernel function, $\alpha_i$ and $\alpha_i^*$ are nonzero Lagrangian multipliers and solutions to the dual problem.

B. Long-Short Term Memory (LSTM) Neural Network

LSTM contains special units called memory blocks in the recurrent hidden layer. These memory blocks contain memory cells with self-connections storing the temporal state of the network in addition to special multiplicative units called gates to control the flow of information. Each memory block in the original architecture contained an input gate and an output gate. The input gate controls the flow of input activations into the memory cell. The output gate controls the output flow of cell activations into the rest of the network. Later, the forget gate was added to the memory block [7]. This addressed a weakness of LSTM models preventing them from processing continuous input streams that are not segmented into subsequences. The forget gate scales the internal state of the cell before adding it as an input to the cell through the self-recurrent connection of the cell, therefore adaptively forgetting or resetting the cell's memory. In addition, the modern LSTM architecture contains peephole connections from its internal cells to the gates in the same cell to learn the precise timing of the outputs [13].

An LSTM network maps a sequence of input vectors $x = (x_1, ..., x_T)$ to a sequence of output vectors $y = (y_1, ..., y_T)$ by iteratively calculating the network unit activations using the following equations from $t = 1 ... T$:

$$\text{Block input: } z_t = g(W_z x_t + V_z y_{t-1} + b_t) \quad (5)$$

$$\text{Input gate: } i_t = \sigma(W_i x_t + V_i y_{t-1} + b_i) \quad (6)$$

$$\text{Forget gate: } f_t = \sigma(W_f x_t + V_f y_{t-1} + b_f) \quad (7)$$

$$\text{Memory cell: } c_t = i_t \odot z_t + f_t \odot c_{t-1} \quad (8)$$

$$\text{Output gate: } o_t = \sigma(W_o x_t + V_o y_{t-1} + b_0) \quad (9)$$

$$\text{Block output: } y_t = o_t \odot g(c_t) \quad (10)$$

where $W$ and $V$ denote input and recurrent weight matrices, respectively (e.g. $W_i$ is the weight matrix from the current input step to input gate $i$; and $V_i$ is the recurrent weight matrix from the output of the previous step to input gate $i$), the $b$ terms denote bias weight vectors (e.g. $b_i$ is the input gate bias weight vector), $\sigma$ is the logistic sigmoid function, $g(x) = tanh(x)$ and $i, f, o$ and $c$ are the input gate, forget gate, output gate and memory cell activation vectors, respectively, all of which are the same size as the cell output activation vector m, $\odot$ is the element-wise product of the vectors.

C. M-Band Wavelets

An M-Band Wavelet Transform [14][15][21][25][26][31] is totally determined by M sets of filter banks with certain properties. In our research, we used a 4-Band Wavelet Transform with the following filter banks:

α= [-0.06737176, 0.09419511, 0.40580489, 0.56737176, 0.56737176, 0.40580489, 0.09419511, -0.06737176]
β= [-0.09419511, 0.06737176, 0.56737176, 0.40580489, -0.40580489, -0.56737176, -0.06737176, 0.0941951]
γ= [-0.09419511, -0.06737176, 0.5673717, -0.4058048, -0.4058048, -0.5673717, -0.06737176, -0.0941951]
δ= [-0.06737176, -0.09419511, 0.40580489, -0.56737176, 0.56737176, -0.40580489, 0.09419511, 0.06737176].

Amongst the filter banks, α is the low pass filter bank, while β, γ, and δ are high pass filter banks. These filter banks satisfy the following conditions:

$$\sum_{i=1}^{8} \alpha_i = 2 \tag{11}$$

$$\sum_{i=1}^{8} \beta_i = \sum_{i=1}^{8} \gamma_i = \sum_{i=1}^{8} \delta_1 = 0 \tag{12}$$

$$|\alpha| = |\beta| = |\gamma| = |\delta| \tag{13}$$

$$\alpha \cdot \beta = \alpha \cdot \gamma = \alpha \cdot \delta = \beta \cdot \gamma = \beta \cdot \delta = \gamma \cdot \delta = 0 \tag{14}$$

If we are working with the signal $S \in \mathbb{R}^{4^k} (k \in \mathbb{N}, k \geq 2)$, we can create a corresponding $4^k \times 4^k$ Wavelet Transform matrix $T_1$ by shifting and wrapping around the filter banks α, β, γ, and δ as shown in Figure 1. Then the first level Wavelet Transform of $S$ can be done by:

$$T_1 S = [a^1 \quad d_1^1 \quad d_1^2 \quad d_1^3]^T \triangleq \tilde{S}_1 \tag{15}$$

Where:
$$a^1 = [a_1^1, a_2^1, a_{13}^1, \ldots, a_{4^{k-1}}^1] \tag{16}$$

$$d_i^1 = [d_{i1}^1, d_{i2}^1, d_{i3}^1, \ldots, d_{i,4^{k-1}}^1], i = 1,2,3 \tag{17}$$

$$\begin{bmatrix}
\alpha_1 & \alpha_2 & \alpha_3 & \alpha_4 & \alpha_5 & \alpha_6 & \alpha_7 & \alpha_8 & 0 & 0 & 0 & 0 & 0 & 0 & 0 & 0 \\
0 & 0 & 0 & 0 & \alpha_1 & \alpha_2 & \alpha_3 & \alpha_4 & \alpha_5 & \alpha_6 & \alpha_7 & \alpha_8 & 0 & 0 & 0 & 0 \\
0 & 0 & 0 & 0 & 0 & 0 & 0 & 0 & \alpha_1 & \alpha_2 & \alpha_3 & \alpha_4 & \alpha_5 & \alpha_6 & \alpha_7 & \alpha_8 \\
\alpha_5 & \alpha_6 & \alpha_7 & \alpha_8 & 0 & 0 & 0 & 0 & 0 & 0 & 0 & 0 & \alpha_1 & \alpha_2 & \alpha_3 & \alpha_4 \\
\beta_1 & \beta_2 & \beta_3 & \beta_4 & \beta_5 & \beta_6 & \beta_7 & \beta_8 & 0 & 0 & 0 & 0 & 0 & 0 & 0 & 0 \\
0 & 0 & 0 & 0 & \beta_1 & \beta_2 & \beta_3 & \beta_4 & \beta_5 & \beta_6 & \beta_7 & \beta_8 & 0 & 0 & 0 & 0 \\
0 & 0 & 0 & 0 & 0 & 0 & 0 & 0 & \beta_1 & \beta_2 & \beta_3 & \beta_4 & \beta_5 & \beta_6 & \beta_7 & \beta_8 \\
\beta_5 & \beta_6 & \beta_7 & \beta_8 & 0 & 0 & 0 & 0 & 0 & 0 & 0 & 0 & \beta_1 & \beta_2 & \beta_3 & \beta_4 \\
\gamma_1 & \gamma_2 & \gamma_3 & \gamma_4 & \gamma_5 & \gamma_6 & \gamma_7 & \gamma_8 & 0 & 0 & 0 & 0 & 0 & 0 & 0 & 0 \\
0 & 0 & 0 & 0 & \gamma_1 & \gamma_2 & \gamma_3 & \gamma_4 & \gamma_5 & \gamma_6 & \gamma_7 & \gamma_8 & 0 & 0 & 0 & 0 \\
0 & 0 & 0 & 0 & 0 & 0 & 0 & 0 & \gamma_1 & \gamma_2 & \gamma_3 & \gamma_4 & \gamma_5 & \gamma_6 & \gamma_7 & \gamma_8 \\
\gamma_5 & \gamma_6 & \gamma_7 & \gamma_8 & 0 & 0 & 0 & 0 & 0 & 0 & 0 & 0 & \gamma_1 & \gamma_2 & \gamma_3 & \gamma_4 \\
\delta_1 & \delta_2 & \delta_3 & \delta_4 & \delta_5 & \delta_6 & \delta_7 & \delta_8 & 0 & 0 & 0 & 0 & 0 & 0 & 0 & 0 \\
0 & 0 & 0 & 0 & \delta_1 & \delta_2 & \delta_3 & \delta_4 & \delta_5 & \delta_6 & \delta_7 & \delta_8 & 0 & 0 & 0 & 0 \\
0 & 0 & 0 & 0 & 0 & 0 & 0 & 0 & \delta_1 & \delta_2 & \delta_3 & \delta_4 & \delta_5 & \delta_6 & \delta_7 & \delta_8 \\
\delta_5 & \delta_6 & \delta_7 & \delta_8 & 0 & 0 & 0 & 0 & 0 & 0 & 0 & 0 & \delta_1 & \delta_2 & \delta_3 & \delta_4
\end{bmatrix}$$

Figure 1. An example of $4^2 \times 4^2$ Wavelet Transform matrix

We can verify that $T_1$ is an orthonormal matrix and therefore the column (row) vectors of $T_1$ form a set of an orthonormal basis for $\mathbb{R}^{4^k}$. So $T_1 S$ means we transform S into the corresponding Wavelet domain. Let $C_1, C_2, \ldots, C_{4^k}$ be the column vectors of $T_1^t$, so the transpose of the column vectors of $T_1^t$ are row vectors of $T_1$. Then the components of $\tilde{S}_1$ are coordinates of S under orthonthe ormal basis $\{C_1, C_2, \ldots, C_{4^k}\}$. Hence, the components of $\tilde{S}_1$ are also called the Wavelet coefficients of S. Most importantly, the 4-Band Wavelet Transform of S decomposes S into 4 different frequency components with $a^1$ being the lowest frequency and $d_1^1, d_2^1, d_3^1$ being the higher frequency components of S. If needed, we can apply Wavelet Transformation to $a^1$ using a $4^{k-1} \times 4^{k-1}$ Wavelet Transform matrix $T_2$, such that:

$$T_1 S = [a^1 \quad d_1^1 \quad d_2^1 \quad d_3^1]^T \triangleq \tilde{S}_1 \tag{18}$$

Where:
$$a^2 = [a_1^2, a_2^2, a_3^2, \ldots, a_{4^{k-2}}^2]^T \tag{19}$$

$$d_i^2 = [d_{i1}^2, d_{i2}^2, d_{i3}^2, \ldots, d_{i,4^{k-2}}^2]^T, i = 1,2,3 \tag{20}$$

The 4-Band Wavelet Transform of $a^1$ decomposes $a^1$ into four different frequency components with $a^2$ being lowest frequency and $d_i^2$ (i = 1,2,3) being higher frequency components of $a^1$.

Let:
$$T = \begin{bmatrix} T_2 & 0 \\ 0 & I \end{bmatrix} T_1 \tag{21}$$

where the lower corner 0 is an $3(4^{k-1}) \times (4^{k-1})$ zero matrix, upper corner 0 is a $4^{k-1} \times 3*4^{k-1}$ zero matrix, and I is an $3*4^{k-1} \times 3*4^{k-1}$ identity matrix. Then:

$$TS = [a^2 \quad d_1^2 \quad d_2^2 \quad d_3^2 \quad d_1^1 \quad d_2^1 \quad d_3^1]^T \triangleq \tilde{S}_2 \tag{22}$$

and $\tilde{S}_2$ is the second level Wavelet coordinates of $S$. Since $\{C_1, C_2, \ldots, C_{4^k}\}$ is an orthonormal basis of $\mathbb{R}^{4^k}$,

$$S = s_1 C_1 + s_2 C_2 + \cdots + s_n C_n \tag{23}$$

where $n = 4^k$ and $s_i = C_i^t S = \langle C_i, S \rangle$, the inner product of $C_i$ and $S$ for $i = 1, 2, \ldots, n$. Therefore:

$$s_i = \begin{cases} a_i^1 \text{ for } i = 1,2,\ldots,4^{k-1} \\ d_{1i}^1 \text{ for } i = 4^{k-1}+1, \ldots, 2*4^{k-1} \\ d_{2i}^1 \text{ for } i = 2*4^{k-1}+1, \ldots, 3*4^{k-1} \\ d_{3i}^1 \text{ for } i = 3*4^{k-1}+1, \ldots, 4^k \end{cases} \tag{24}$$

Let:

$$\begin{cases} A^1 = a_1^1 C_1 + \cdots + a_{4^{k-1}}^1 C_{4^{k-1}} \\ D_i^1 = d_{i1}^1 C_{i*4^{k-1}+1} + \cdots + d_{i4^{k-1}}^1 C_{(i+1)*4^{k-1}} \end{cases} \tag{25}$$

for i= 1,2,3. Then $A^1$ is corresponding to $a^1$ and $D_i^1$ is corresponding to $d_i^1$ for i = 1,2,3. Also, let:

$$V_1 = span\{C_1, ..., C_{4^{k-1}}\}$$
$$W_i^1 = span\{C_{i*4^{k-1}+1}, C_{(i+1)*4^{k-1}}\} \quad (26)$$

Then $V_1$, $W_1^1$, $W_2^1$ and $W_3^1$ are orthogonal to each other and the vector space $\mathbb{R}^{4^k}$ is the direct sum of $V_1$, $W_1^1$, $W_2^1$ and $W_3^1$, that is:

$$\mathbb{R}^{4^k} = V_1 \oplus W_1^1 \oplus W_2^1 \oplus W_3^1 \quad (27)$$

Moreover, we have:

$$S = A^1 + D_1^1 + D_2^1 + D_3^1 \quad (28)$$

where $A^1$ is the orthogonal projection of $S$ onto $V_1$, $D_i^1$ is orthogonal projection of S onto $W_i^1$, for i = 1,2,3. If we apply the second level Wavelet Transform, then:

$$A^1 = A^2 + D_1^2 + D_2^2 + D_3^2$$
$$S = A^2 + D_1^2 + D_2^2 + D_3^2 + D_1^1 + D_2^1 + D_3^1 \quad (29)$$

IV. EXPERIMENT PROCEDURE AND RESULTS

A. Wavelet Denoising

In this research, a discrete 4-Band 2 regular Wavelet Transform was used, such that:

$$T * S = \begin{bmatrix} A \\ d_1 \\ d_2 \\ d_3 \end{bmatrix} = \tilde{S} \quad (30)$$

$$A = \begin{bmatrix} <T_1, S> \\ <T_2, S> \\ \vdots \\ <T_{\frac{n}{4}-1}, S> \\ <T_{\frac{n}{4}}, S> \end{bmatrix} \quad (31)$$

$$\begin{bmatrix} d_1 \\ d_2 \\ d_3 \end{bmatrix} = \begin{bmatrix} <T_{\frac{n}{4}+1}, S> \\ <T_{\frac{n}{4}+2}, S> \\ \vdots \\ <T_n, S> \end{bmatrix} \quad (32)$$

$T$ = Wavelet Transform matrix, $T^T = T^{-1}$, the transpose of $T$, $S$ = Historical stock prices, $A$ = Approximation portion (low frequency), $d_i$ = Detail portions ($i$ = 1, 2,3), $n$ = Number of days (must be in the form $n = 4^k$).
In this research, a sample consisting of 256 trading days was used, and thus k = 3, and the resulting Wavelet Matrix was of size 256x 256.

The following threshold was then added to prevent over-smoothening of data [14][31]:

$$\lambda = \sigma\sqrt{2\log(N)} \quad (33)$$

$\sigma$ = Standard deviation of noise from Wavelet coefficients
$N$ = Number of elements in the band
The dataset's standard deviation cannot be directly calculated; however, it can be estimated based on the following [14]:

$$\sigma_i \approx \frac{Median(|d_{ij}|)}{0.6745} \quad (34)$$

Thresholding can now be done by redefining $d_i^j$ as follows:

$$d_i^j = \begin{cases} 0, & |d_i^j| < \lambda_i \\ d_i^j, & |d_i^j| \geq \lambda_i \end{cases} \quad (35)$$

After thresholding, a denoised version of $\tilde{S}$ is obtained, which we denote as $\hat{S}$. This data must now be reconstructed out of the Wavelet domain. This can be done as follows:

$$T^T \cdot \hat{S} = S^* \approx S \quad (36)$$

The data is now ready to be used. The example below showcases the smoothening effect the Wavelet Transform on a dataset. In this example, 256 trading worth of Apple stock was used.

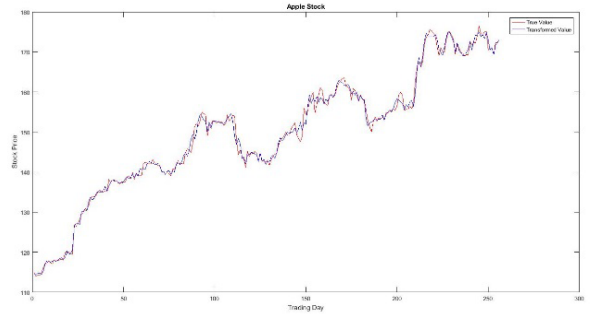

Figure 2. Smoothened Apple Stock: 12/2/2016 – 01/04/2018

One can observe that the Wavelet Denoised stocks are smoother when compared to the original, but not excessively so.
Normally an analyst would use historical stock data and find the weighted moving average associated with it. They would then use this data to obtain other stock technical indicators such as Welles Wilder's Directional Movement Indicator, Commodity Channel Index, Momentum Oscillator, Price Oscillator, Rate of Change, Relative Strength Indicator…etc., for stock prediction algorithms.
In this research, Wavelet Denoised stock data was used for training various machine learning methods (namely, SVR and NN), and its effectiveness was then analyzed.

B. LSTM Neural Network

In this research, a Nonlinear Autoregressive with External (Exogenous) Input (NARX), and a Levenberg-

Marquardt (i.e. Dampened Least-Squares) training algorithm was used.

Utilizing MATLAB's Neural Net Time Series Toolbox, a 10 hidden layers neural net was created to predict Apple stock using historical data from Apple, as well as 23 other relevant stocks:

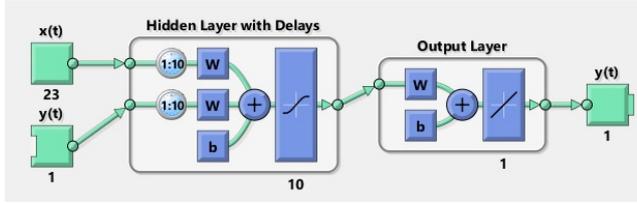

Figure 3. 10 layers Neural Net model.

The model was trained by the neural network using historical data as an input and then used to predict the stock's value for the following day. The table below features a comparison of RMSE values for 1-day-historical, 5-day-historical and 10-day-historical values for Apple Stock, both with and without the Wavelet Denoising.

TABLE I. RMSE COMPARISION FOR NEURAL NETWORK MODELS

|  | Number of historical data | | |
|---|---|---|---|
|  | *1 Day* | *5 Days* | *10 Days* |
| Original Data | 10.098 | 5.68 | 3.416 |
| Wavelet Denoised Data | 5.254 | 1.194 | 0.879 |

From the table above, we can observe that an increase in historical data leads to lower prediction error. The Wavelet Denoising was also shown to improve prediction rates compared to untreated data.

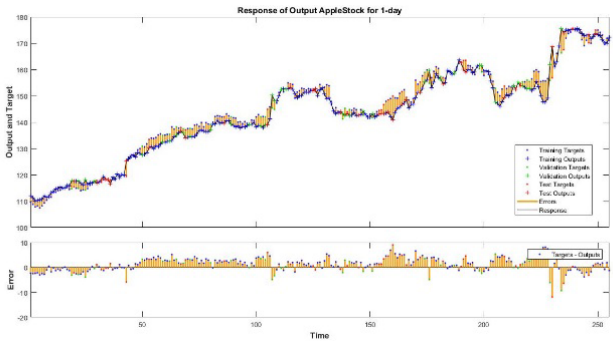

Figure 4. NN Model for Apple Stock using 1-day original historical data

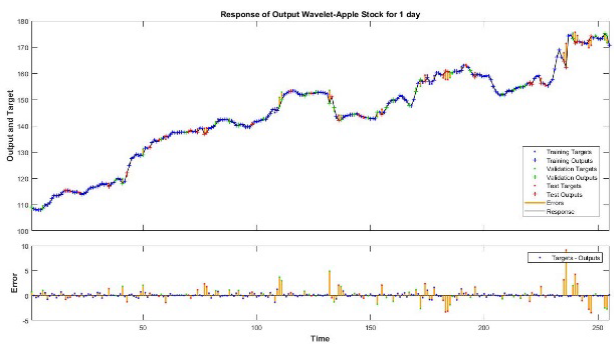

Figure 5. NN Model for Apple Stock using 1-day denoised historical data.

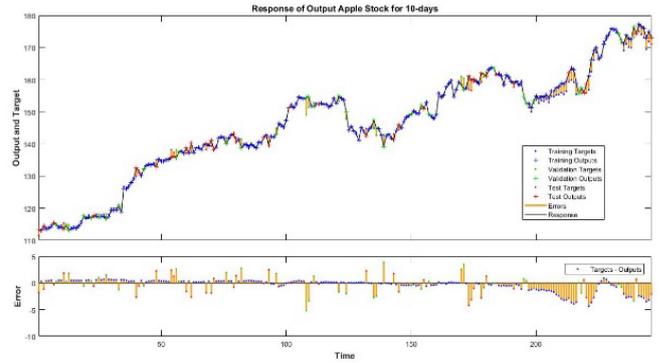

Figure 6. NN Model for Apple Stock using 10-day original historical data

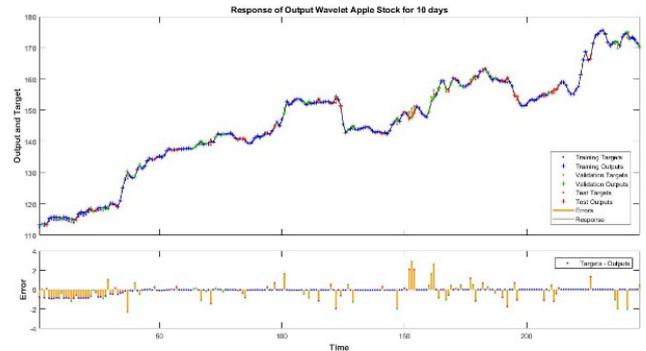

Figure 7. NN Model for Apple Stock using 10-day denoised historical data

As can be seen in the above figures, the training error is significantly less in models pretreated with Wavelet Denoising, and thus prediction accuracy was increased.

## C. Support Vector Regression

Utilizing MATLAB's Regression Toolbox, various SVR models were created using different kernels. Historical stock data from Apple, as well as 23 other relevant stocks, were used in training an SVR model so that it uses the data from one day to predict the stock's movement on the next day, 5 days, 10 days, and so on in a recursive manner. Examples from this method can be found below:

TABLE II. COMPARISION BETWEEN DIFFERENT SVR KEVEL FOR APPLE STOCK NEXT DAY PREDICTION

| Kernel | Error and best-fit | | | | | |
|---|---|---|---|---|---|---|
|  | *R-Squared* | | *RMSE* | | *MAE* | |
|  | *Original* | *Wavelet* | *Original* | *Wavelet* | *Original* | *Wavelet* |
| Linear | 0.98 | 0.97 | 1.9631 | 1.9893 | 1.4673 | 1.4586 |
| Quadratic | 0.97 | 0.97 | 2.2321 | 2.1907 | 1.6951 | 1.6125 |
| Cubic | 0.95 | 0.96 | 2.5139 | 2.237 | 1.8821 | 1.8031 |
| Fine Gaussian | 0.94 | 0.95 | 3.1253 | 2.7608 | 2.3987 | 2.1611 |
| Medium Gaussian | 0.97 | 0.97 | 2.002 | 1.9795 | 1.54 | 1.5219 |
| Coarse Gaussian | 0.94 | 0.95 | 2.9331 | 2.883 | 2.2317 | 2.2359 |

TABLE III. COMPARISION BETWEEN DIFFERENT SVR KEVEL FOR APPLE STOCK NEXT 5-DAY PREDICTION

| Kernel | Error and best-fit | | | | | |
|---|---|---|---|---|---|---|
| | R-Squared | | RMSE | | MAE | |
| | Original | Wavelet | Original | Wavelet | Original | Wavelet |
| Linear | 0.93 | 0.93 | 3.1369 | 3.1245 | 2.4152 | 2.2966 |
| Quadratic | 0.97 | 0.97 | 2.2165 | 2.2238 | 1.17402 | 1.7443 |
| Cubic | 0.92 | 0.97 | 3.431 | 2.2292 | 2.1515 | 1.7323 |
| Fine Gaussian | 0.93 | 0.94 | 3.1304 | 2.9433 | 2.2481 | 2.2188 |
| Medium Gaussian | 0.96 | 0.96 | 2.3102 | 2.3649 | 1.7631 | 1.8457 |
| Coarse Gaussian | 0.91 | 0.91 | 3.6734 | 3.7127 | 2.912 | 2.951 |

TABLE IV. COMPARISION BETWEEN DIFFERENT SVR KEVEL FOR APPLE STOCK NEXT 13-DAY PREDICTION

| Kernel | Error and best-fit | | | | | |
|---|---|---|---|---|---|---|
| | R-Squared | | RMSE | | MAE | |
| | Original | Wavelet | Original | Wavelet | Original | Wavelet |
| Linear | 0.92 | 0.93 | 3.371 | 3.2017 | 2.4475 | 2.3962 |
| Quadratic | 0.95 | 0.97 | 2.7389 | 2.1604 | 1.9181 | 1.67 |
| Cubic | 0.94 | 0.94 | 2.8358 | 2.2697 | 2.0894 | 1.765 |
| Fine Gaussian | 0.92 | 0.94 | 3.3907 | 2.9749 | 2.5148 | 2.1824 |
| Medium Gaussian | 0.96 | 0.94 | 2.4995 | 2.3642 | 1.9466 | 1.8496 |
| Coarse Gaussian | 0.85 | 0.85 | 4.7299 | 4.6785 | 3.7144 | 3.6206 |

TABLE V. COMPARISION BETWEEN DIFFERENT SVR KEVEL FOR APPLE STOCK NEXT 21-DAY PREDICTION

| Kernel | Error and best-fit | | | | | |
|---|---|---|---|---|---|---|
| | R-Squared | | RMSE | | MAE | |
| | Original | Wavelet | Original | Wavelet | Original | Wavelet |
| Linear | 0.89 | 0.89 | 3.8937 | 3.8688 | 2.9954 | 2.9182 |
| Quadratic | 0.96 | 0.96 | 2.4058 | 2.4013 | 1.8694 | 1.8403 |
| Cubic | 0.96 | 0.96 | 2.404 | 2.3538 | 1.9004 | 1.8131 |
| Fine Gaussian | 0.94 | 0.95 | 2.9382 | 2.633 | 2.1888 | 1.9919 |
| Medium Gaussian | 0.95 | 0.95 | 2.729 | 2.6223 | 2.058 | 1.9545 |
| Coarse Gaussian | 0.82 | 0.82 | 4.9184 | 4.9072 | 3.9097 | 3.9166 |

Most of the resulting models have a strong fit, with relatively high $R^2$ values. These results also suggest that the Wavelet Transform successfully smoothens data and leads to increased prediction accuracy, both for SVR and NN models. Both machine learning methods, SVR and NN, also proved effective on their own. Within SVR, certain kernels, such as the cubic kernel, had relatively low RMSE values though they performed poorly with regards to stock forecasting. In this research, linear and Gaussian kernels were found to be most accurate [30].

V. DISCUSSION AND CONCLUSION

The results of this research indicate the suggested approach being an efficient method with regards to stock forecasting. The Wavelet Transform and denoising technique can be used to handle stock data, and then be used to more efficiently train machine learning models. Both SVR and NN models were found to be accurate machine learning methods despite their different approaches; recursively using previously predicted data to predict the following day for long-term forecasting.

Forecasting rates were also found to be more accurate when predicting $F$-number of days, where $F$ is a Fibonacci number. Although interesting, this phenomenon is in line with current prediction methods used by those in the financial sector.

The results of this research can be extended by combining both machine learning techniques into one overlying model (Wavelet-based SVR-NN), rather than having them as separate approaches. We also wish to further explore various other kernels within SVR, namely the newly established Wavelet kernel and gauge its performance with regards to forecasting accuracy. We believe the results of this research can also be improved better selecting the basket of stocks used to train our machine learning models.

ACKNOWLEDGMENT

We would like to thank Western Connecticut State University, WCSU Student Government Association, WCSU Kathwari Honors Program, and Mr. Sherman Tao for their continued financial support.